\documentclass[aps,prl,twocolumn]{revtex4-2}

\usepackage{amsmath}
\usepackage{array}
\usepackage{amsthm}
\usepackage{amssymb}
\usepackage{bbm}
\usepackage{mathrsfs}
\usepackage{indentfirst}
\usepackage{mathtools}
\usepackage{float}
\usepackage{hyperref}
\usepackage{graphicx}
\usepackage{placeins}
\usepackage{subcaption}
\usepackage{color}

\newcommand{\PP}{\mathcal{P}}

\newcommand{\dd}{\mathrm{d}}

\newcommand{\xx}{\tilde{x}}
\newcommand{\XX}{\widetilde{X}}

\newcommand*\sca[1]{\left\langle#1\right\rangle}

\begin{document}

\title{Fluctuation Theorems and Thermodynamic Inequalities for Nonequilibrium Processes Stopped at Stochastic Times}

\author{Haoran Yang\textsuperscript{1,3}}
\email{yanghr@pku.edu.cn}

\author{Hao Ge\textsuperscript{1,2}}
\email{haoge@pku.edu.cn}

\affiliation{
	\textsuperscript{1}Beijing International Center for Mathematical Research (BICMR), Peking University, Beijing 100871, People's Republic of China \\
	\textsuperscript{2}Biomedical Pioneering Innovation Center (BIOPIC), Peking University, Beijing 100871, People's Republic of China \\
	\textsuperscript{3}School of Mathematical Sciences, Peking University, Beijing 100871, People's Republic of China
}

\date{\today}

\begin{abstract}
We investigate thermodynamics of general nonequilibrium processes stopped at stochastic times. We propose a systematic strategy for constructing fluctuation-theorem-like martingales for each thermodynamic functional, yielding a family of stopping-time fluctuation theorems. We derive second-law-like thermodynamic inequalities for the mean thermodynamic functional at stochastic stopping times, the bounds of which are stronger than the thermodynamic inequalities resulting from the traditional fluctuation theorems when the stopping time is reduced to a deterministic one. Numerical verification is carried out for three well-known thermodynamic functionals, namely, entropy production, free energy dissipation and dissipative work. These universal equalities and inequalities are valid for arbitrary stopping strategies, and thus provide a comprehensive framework with new insights into the fundamental principles governing nonequilibrium systems.
\end{abstract}

\maketitle


Stochastic thermodynamics extends classical thermodynamics to individual trajectories of non-equilibrium processes, encompassing stationary or transient systems with or without external driving forces \cite{Sekimoto1998,seifert2012stochastic,Pigolotti2017,Seifert2008}. A first-law-like energy balance equality and various second-law-like thermodynamic inequalities can be derived from fluctuating trajectories. Fluctuation theorems emerging from stochastic thermodynamics, as equality versions of the second law, impose constraints on probability distributions of thermodynamic functionals along single stochastic trajectories  \cite{jarzynski1997nonequilibrium,Jarzynski2011,crooks1999entropy,Seifert2005,collin2005verification,esposito2010letter,Qian2020,Manzano2015,Lahiri2014,Chetrite2011,Crooks2000,Chernyak2006,Speck2005,Harris2007,Evans2002}.

Recently, a gambling demon, which stops the processes at random times, has been proposed for non-stationary stochastic processes without external driving force and feedback of control under an arbitrary deterministic protocol \cite{Manzano2021,roldan2023martingales}. The demon employs  martingales, a concept that has been proposed in probability theory for more than $70$ years. The authors constructed a martingale for dissipative work, and obtained a stopping-time fluctuation theorem by applying the well-known optional stopping theorem (or Doob's optional sampling theorem), which states that the average of a martingale at a stopping time is equal to the average of its initial value \cite{Williams1991}. 

On the other hand, we already know that there are three faces in stochastic thermodynamics \cite{Ge2010,esposito2010letter,esposito2010three,Ge2009}, namely, (total) entropy production, housekeeping heat (non-adiabatic entropy production) and free energy dissipation (adiabatic entropy production). In a system with no external driving force, the housekeeping heat vanishes and the entropy production is equal to the free energy dissipation. However, in general non-stationary stochastic processes with an external driving force as well as an time-dependent protocol, we are curious about whether different martingales can be constructed for entropy production and free energy dissipation separately, while the martingale for housekeeping heat is straightforward to construct without any compensated term \cite{chetrite2019martingale}. Both entropy production and free energy dissipation belong to a class of functionals along a single stochastic trajectory, i.e. general backward thermodynamic functionals, which has been rigorously defined in \cite{Ge2021}. Housekeeping heat belongs to another class, called forward thermodynamic functionals \cite{Ge2021}.

Therefore, in this paper, we propose a systematic strategy for constructing martingales applicable to general backward thermodynamic functionals, with a focus on entropy production, free energy dissipation, and dissipative work as illustrative examples. Notably, the construction of martingales for forward thermodynamic functionals has been previously established in \cite{Ge2021}. By leveraging our constructed martingales, we derive stopping-time fluctuation theorems that hold for general backward thermodynamic functionals, followed by second-law-like thermodynamic inequalities for arbitrary stopping times.
When the stochastic stopping time reduces to a deterministic one, we exploit the additional degree of freedom present in our constructed martingales, enabling us to obtain a sharper nonnegative bound for the mean thermodynamic functional. In particular, we obtain a stronger inequality for the dissipative work than that obtained through classic Jarzynski equality.

{\em Stopping-time fluctuation theorems and thermodynamic inequalities} First, we will give an even more general definition of the backward thermodynamic functional than \cite{Ge2021}.
We consider a stochastic thermodynamic system with temperature $\beta=\frac{1}{k_B\mathbf{T}}$. We denote the state (discrete or continuous) of the system at time $s \geqslant 0$ by $X(s)$, whose stochastic dynamics is governed by a prescribed deterministic protocol $\Lambda=\{\lambda(s) \colon s\geqslant 0\}$. For a given duration $[0, t]$, the trajectories are traced by the coordinates in phase space, denoted by $x_{[0, t]} \equiv \{ x(s) \}_{0 \leqslant s \leqslant t}$. We further denote the probability of observing a given trajectory $x_{[0, t]}$ by $\PP^X (x_{[0, t]})$, and the probability density of $X(s)$ by $\varrho^X(x,s)$ at any given time $s$. The general backward thermodynamic functional in the duration $[0,t]$ is defined by $\{X(s)\}_{0\leqslant s\leqslant t}$ and another stochastic process $\{Y(s)\}$ with protocol $\tilde{\Lambda}=\{\tilde{\lambda}(s) \colon s\geqslant 0\}$ (can be either the same as or different from $\Lambda$). The only condition is that the processes $\{X(s)\}_{0\leqslant s\leqslant t}$ and $\{Y(s)\}_{0\leqslant s\leqslant t}$ are absolutely continuous with each other, i.e. the probability $\PP^X (x_{[0, t]})>0$ if and only if $\PP^{Y} (x_{[0, t]})>0$ for any given trajectory $x_{[0, t]}$. We define a third process $\{Z^t(s)\}_{0 \leqslant s \leqslant t}$ driven by the time-reversed protocol $\tilde{\Lambda}^{r,t}=\{\tilde{\lambda}(t-s) \colon 0\leqslant s\leqslant t\}$ of $\{Y(s)\}$ up to time $t$. The probability density of $Z^t(s)$ is denoted by $\varrho^{Z^t}(x,s)$ for any given time $s\leqslant t$. Note that there is also an additional degree of freedom, i.e. the arbitrary choice of the initial distribution $\varrho^{Z^t}(x,0)$ of $\{Z^t(s)\}_{0\leqslant s\leqslant t}$ for any $t$, because for different $t$, only the protocols inherited from $\{Y(s)\}$ are closely related to each other, not the initial distributions.

The probability of observing a given trajectory $x_{[0,t]}$ in $\{Z^t(s)\}_{0\leqslant s\leqslant t}$ is denoted by $\PP^{Z^t}(x_{[0,t]})$. We define a general backward thermodynamic functional by
\begin{equation*}
	F_t(x_{[0, t]}) \equiv \frac{1}{\beta}\ln \frac{ \PP^X (x_{[0, t]}) }{ \PP^{Z^t} (\xx_{[0, t]}) },
\end{equation*}
where $\xx_{[0, t]} \equiv \{ x (t - s) \}_{0 \leqslant s \leqslant t}$ denotes the time reversal of $x_{[0, t]}$ in the duration $[0,t]$.

It is straightforward to derive the fluctuation theorem for $F_t$:
\begin{equation*}
	\sca{ e^{-\beta F_t} } = 1.
\end{equation*}
However, $F_t$ is generally not a martingale \cite{Ge2021}.

For any given time interval $[0,T]$, we would like to add a compensated term $\delta_t$ as a function of $X(t)$ and $t$, to $F_t$, so that $e^{-\beta (F_t+\delta_t)}$ be a martingale, i.e.
$$\sca{e^{-\beta (F_T+\delta_T)} \middle\vert X_{[0,t]}}=e^{-\beta (F_t+\delta_t)},$$
for any $t\in [0,T]$.

Then we propose
\begin{equation}\label{delta}
	\delta_t(X(t)) \equiv \frac{1}{\beta}\ln \frac{\varrho^{Z^t} (X(t), 0)}{\tilde{\varrho}^{Z^T} (X(t), T - t )},
\end{equation}
in which $\tilde{\varrho}^{Z^T} (X(t), T - t )$ is the distribution of $Z^T(t)$ with any arbitrary initial distribution $\tilde{\varrho}^{Z^T} (\cdot,0)$, which is not necessarily the same as  $\varrho^{Z^T} (\cdot, 0)$ and contributes another extra degree of freedom.

We apply the optional stopping theorem in martingale theory to derive the general stopping-time fluctuation theorem
\begin{equation}\label{eq:thm}
	\sca{ e^{-\beta (F_\tau+\delta_\tau)} } = 1,
\end{equation}
where $\tau$ is a stopping time, defined by any stopping mechanism to decide whether to stop a process based on the current position and past events.

By Jensen's inequality,
\begin{equation} \label{ineq1}
	\sca{ F_\tau } \geqslant -\sca{ \delta_\tau }.
\end{equation}
The left-hand-side is independent of $\tilde{\varrho}^{Z^T}$. Hence we can improve the above inequality into
\begin{equation} \label{ineq2}
	\sca{ F_\tau } \geqslant \sup_{\tilde{\varrho}^{Z^T}} -\sca{ \delta_t }.
\end{equation}

A special situation is when $\tau=t$ with probability $1$, followed by
\begin{equation*}\label{eq:thm_cons}
	\sca{ e^{-\beta (F_t+\delta_t)} } = 1,
\end{equation*}
and
\begin{equation}\label{ineq2_cons}
	\sca{ F_t } \geqslant \sup_{\tilde{\varrho}^{Z^T} }-\sca{ \delta_t }= \frac{1}{\beta} \sca{\ln \frac{\varrho^X (X(t), t)}{\varrho^{Z^t} (X(t), 0)}}\geqslant 0,
\end{equation}
in which $\sca{\ln \frac{\varrho^X (X(t), t)}{\varrho^{Z^t} (X(t), 0)}}$ is the relative entropy of $\varrho^X (\cdot, t)$ with respect to $\varrho^{Z^t} (\cdot, 0)$. The inequality (\ref{ineq2_cons}) is stronger than the traditional inequality $\sca{ F_t } \geqslant 0$ derived from the well-known fluctuation theorem $\sca{ e^{-\beta F_t}}=1$, as long as  $\varrho^X (\cdot, t)$ is different from $\varrho^{Z^t} (\cdot, 0)$.

As a corollary, we can derive certain bound for the infimum of $F_t+ \delta_t$ following the strategy in \cite{neri2017statistics}, which holds for both equilibrium processes and general nonequilibrium processes. According to Doob's maximal inequality, we find the following bound for the cumulative distribution of the supremum of $e^{- \beta(F_t + \delta_t)}$,
\begin{equation*}
	\Pr \left( \sup_{0 \leqslant t \leqslant T} e^{- \beta(F_t+\delta_t)} \geqslant \lambda \right) \leqslant \frac{1}{\lambda} \sca{e^{- \beta(F_t+\delta_t)}} = \frac{1}{\lambda},
\end{equation*}
for any $\lambda \geqslant 0$. It is equivalent to a lower bound on the cumulative distribution of the infimum of $\beta(F_t+\delta_t)$ in the given duration $[0, T]$, i.e.
\begin{equation*}
	\Pr \left( \inf_{0 \leqslant t \leqslant T} \{ \beta(F_t+\delta_t) \} \geqslant -s \right) \geqslant 1 -e^{-s},
\end{equation*}
for $s \geqslant 0$. It implies the random variable $-\inf_{0 \leqslant t \leqslant T} \{ \beta(F_t+\delta_t) \}$ dominates stochastically over an exponential random variable with the mean of $1$. Thus, we find the following universal bound for the mean infimum of $\beta(F_t+\delta_t)$, i.e.
\begin{equation*}
	\sca{\inf_{0 \leqslant t \leqslant T} \{(F_t+\delta_t) \}} \geqslant -\frac{1}{\beta}=-k_B\mathbf{T}.
\end{equation*}


{\em Applications} The thermodynamic functional $F_t$ becomes the (total) entropy production $S_\text{tot} (t)$ up to time $t$ if the process $\{Y(t)\}$ is driven by exactly the same protocol as $\{X(t)\}$, and the initial distribution of $Z^t$ is taken to be the distribution of $X(t)$ \cite{Jiang2004,Seifert2005}, i.e. $\varrho^{Z^t}(x,0)=\varrho^X(x,t)$. Then
\begin{equation}
	\delta^{S_\text{tot}}_t(X(t)) \equiv \frac{1}{\beta}\ln \frac{\varrho^X (X(t), t)}{\tilde{\varrho}^{Z^T} (X(t), T - t )},
\end{equation}
and $e^{-\beta (S_\text{tot}(t)+\delta^{S_\text{tot}}_t)}$ is a martingale. It is followed by
\begin{equation}\label{eq:thm_Stot}
	\sca{ e^{-\beta (S_\text{tot} (\tau)+\delta^{S_\text{tot}}_\tau)} } = 1,
\end{equation}
for any stopping time $\tau$, and $\sca{S_\text{tot} (\tau) } \geqslant -\sca{ \delta^{S_\text{tot}}_\tau }$.


The thermodynamic functional $F_t$ becomes the free energy dissipation (adiabatic entropy production) $f_d (t)$  if the process $\{Y(t)\}$ is taken to be the adjoint process of $\{X(t)\}$, and also the initial distribution of $Z^t$ is set as the distribution of $X(t)$, i.e. $\varrho^{Z^t}(x,0)=\varrho^X(x,t)$ \cite{esposito2010letter,esposito2010three,Ge2009,Ge2010}. Then
\begin{equation}
	\delta^{f_d}_t(X(t))\equiv \frac{1}{\beta}\ln \frac{\varrho^X (X(t), t)}{\tilde{\varrho}^{Z^T} (X(t), T - t )},
\end{equation}
and $e^{-\beta (f_d (t)+\delta^{f_d}_t)}$ is a martingale. It is followed by
\begin{equation}\label{eq:thm_fd}
	\sca{ e^{-\beta (f_d(\tau)+\delta^{f_d}_\tau)} } = 1,
\end{equation}
for any stopping time $\tau$, and $\sca{f_d(\tau) } \geqslant -\sca{ \delta^{f_d}_\tau }$.

Let $\pi^X(t)$ be the pseudo-stationary distribution of $X(t)$ corresponding to the protocol $\lambda(t)$, i.e. the stationary distribution of $\{X(t)\}$ if the protocol is fixed at $\lambda(t)$. The thermodynamic functional $F_t$ becomes the dissipative work $W_d (t)$ up to time $t$, if the initial distribution of $X(t)$ is $\pi^X(0)$, the process $\{Y(t)\}$ is taken to be the adjoint process of $\{X(t)\}$, and the initial distribution of $Z^t$ is taken as the pseudo-stationary distribution of $X(t)$, i.e. $\varrho^{Z^t}(x,0)=\pi^X(x,t)$ \cite{Ge2008, Manzano2021}. Then
\begin{equation}
	\delta^{W_d}_t(X(t))\equiv \frac{1}{\beta}\ln \frac{\pi^X (X(t), t)}{\tilde{\varrho}^{Z^T} (X(t), T - t )},
\end{equation}
and $e^{-\beta (W_d (t)+\delta^{W_d}_t)}$ is a martingale. It is followed by
\begin{equation}\label{eq:thm_Wd}
	\sca{ e^{-\beta (W_d(\tau)+\delta^{W_d}_\tau)} } = 1,
\end{equation}
for any stopping time $\tau$, and $\sca{W_d(\tau) } \geqslant -\sca{ \delta^{W_d}_\tau }$.

For the mean $W_d$ up to any fixed time $t$, we can obtain a stronger inequality than $\sca{W_d}\geqslant 0$. Applying (\ref{ineq2_cons}), we have
\begin{equation} \label{ineq2_cons_Wd}
	\sca{W_d(t) } \geqslant \frac{1}{\beta} \sca{\ln \frac{\varrho^X (X(t), t)}{\pi^X (X(t), t)}}\geqslant 0.
\end{equation}
Actually, this inequality can be derived from the equality $\frac{dH(t)}{dt}=-f_d(t)+W_d(t)$ with the inequality $f_d(t)\geqslant 0$ from \cite{Ge2010}, in which $H(t)$ is exactly $\sca{\ln \frac{\varrho^X (X(t), t)}{\pi^X (X(t), t)}}$.
In \cite{Manzano2021}, the thermodynamic functional under investigation is $W_d$ but the $\delta_t$ they defined is the same as $\delta^{S_\text{tot}}_t$. The mathematical derivation here implies that we should use different $\delta_t$ for different thermodynamic functionals.

{\em Numerical verifications} Many mesoscopic biochemical processes such as the kinetics of enzyme or motor molecules, can be modeled in terms of transition rates between discrete states. We apply our theory to a simple stochastic process with only three states. The time-dependent transition rates between different discrete states are set as follows
\begin{eqnarray*}
k_{12}(t)=t,~k_{23}(t)=3t^2,~k_{31}(t)=1;\\
k_{21}=t^2,~k_{32}(t)=2,~k_{13}(t)=2t,
\end{eqnarray*}
in which the chemical driven energy
$$\Delta G(t)=k_BT\ln\frac{k_{12}(t)k_{23}(t)k_{31}(t)}{k_{21}k_{32}(t)k_{13}(t)}=k_BT\ln\frac{3}{4}<0.$$

For the three thermodynamic functionals $S_\text{tot}$, $f_d$ and $W_d$, the stopping strategy for $\tau$ is set as follows: the process is stopped at $\tau < T$ only when the functional reaches a given threshold value before $T$; while the process is stopped at the  final time $\tau = T$ if the threshold value is never reached during the duration $[0, T]$.

\begin{figure*}[htbp]
	\begin{subfigure}[b]{0.32\textwidth}
		\includegraphics[width=\textwidth]{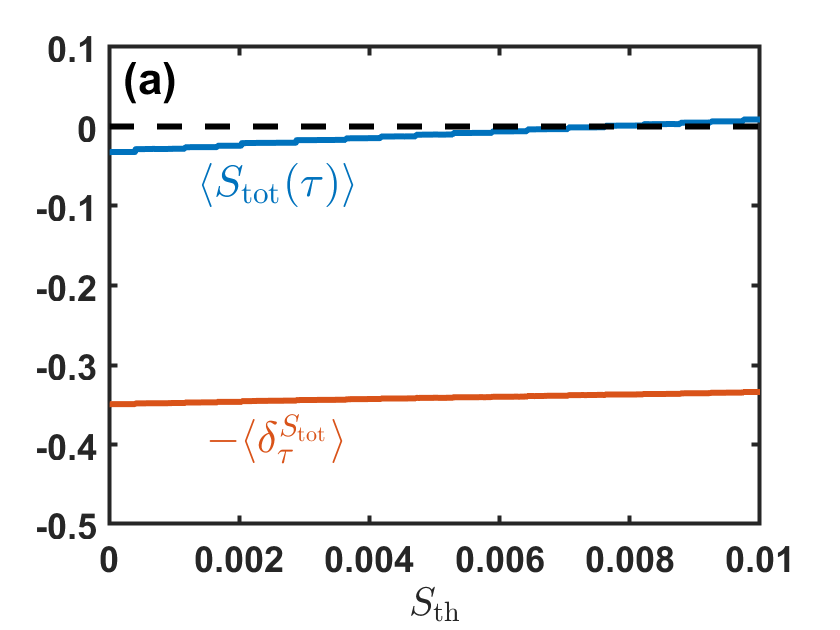}
	\end{subfigure}
	\begin{subfigure}[b]{0.32\textwidth}
		\includegraphics[width=\textwidth]{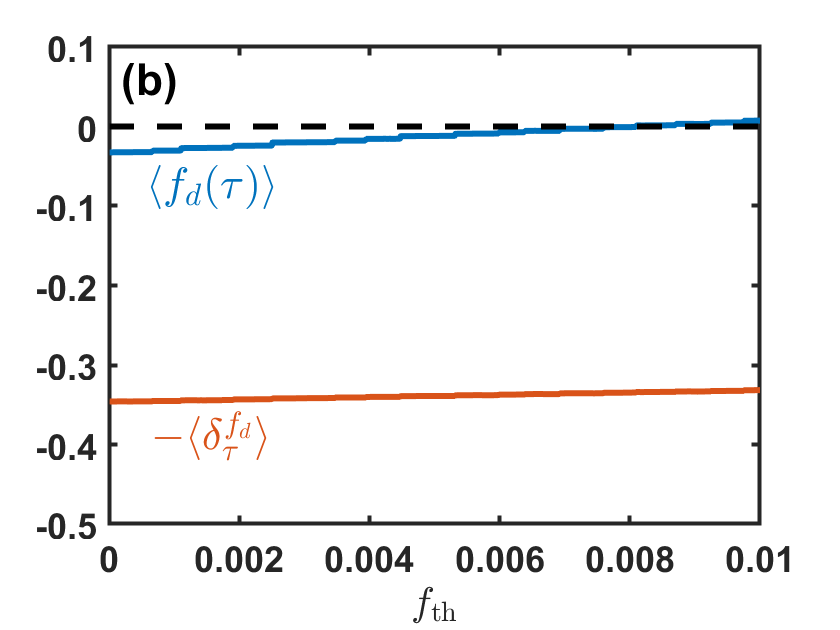}
	\end{subfigure}
	\begin{subfigure}[b]{0.32\textwidth}
		\includegraphics[width=\textwidth]{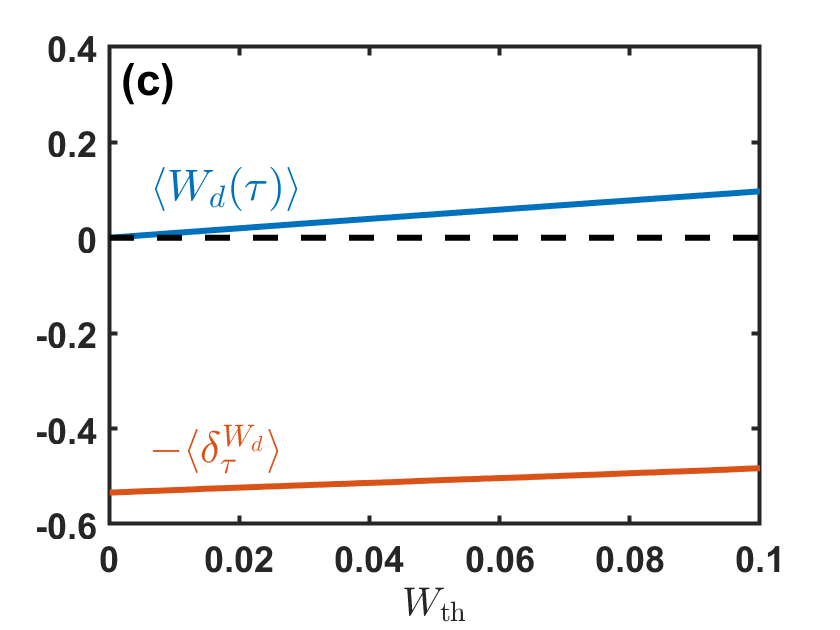}
	\end{subfigure}
	
	\begin{subfigure}[b]{0.32\textwidth}
		\includegraphics[width=\textwidth]{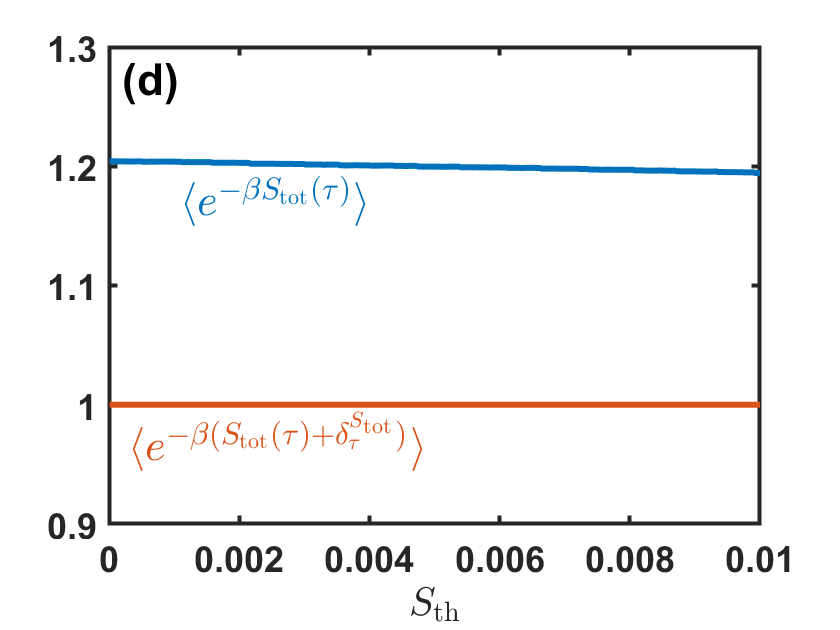}
	\end{subfigure}
	\begin{subfigure}[b]{0.32\textwidth}
		\includegraphics[width=\textwidth]{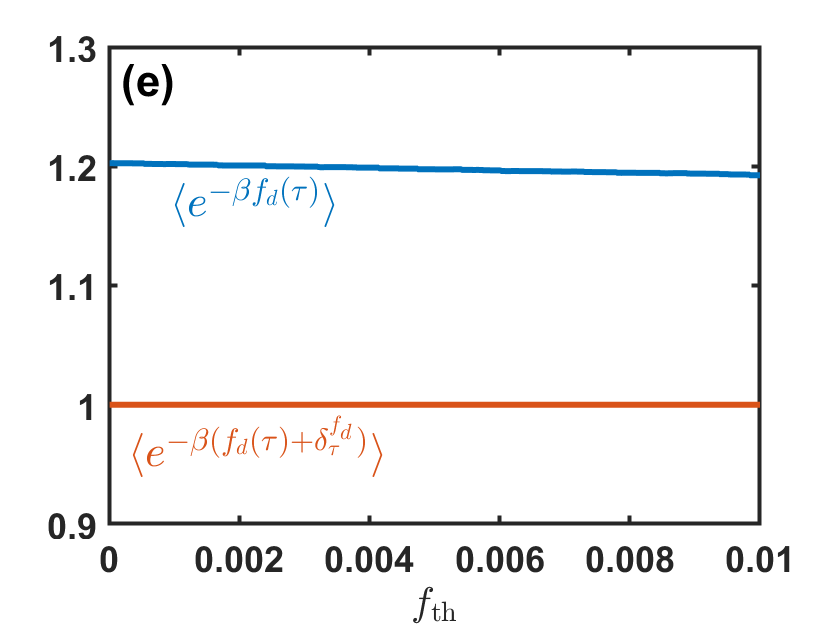}
	\end{subfigure}
	\begin{subfigure}[b]{0.32\textwidth}
		\includegraphics[width=\textwidth]{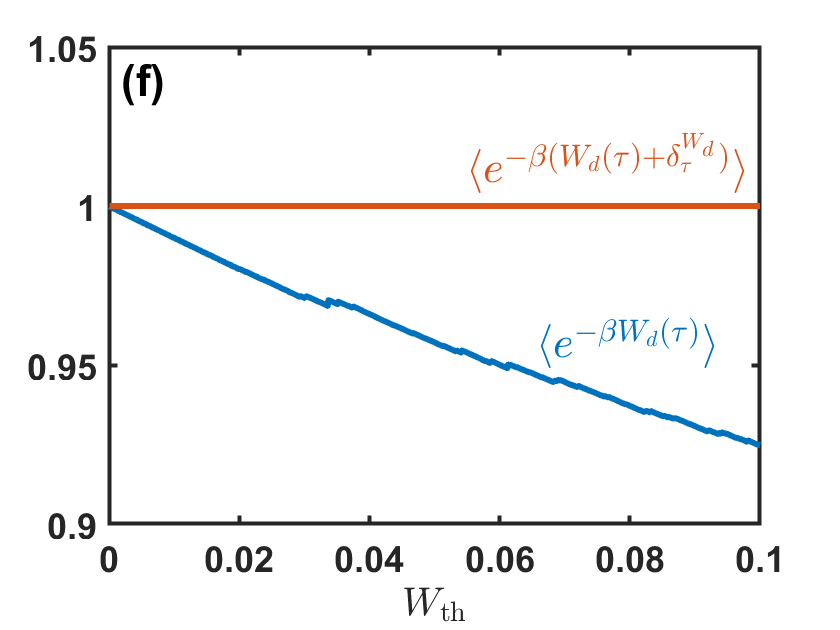}
	\end{subfigure}
	
	\begin{subfigure}[b]{0.4\textwidth}
		\includegraphics[width=\textwidth]{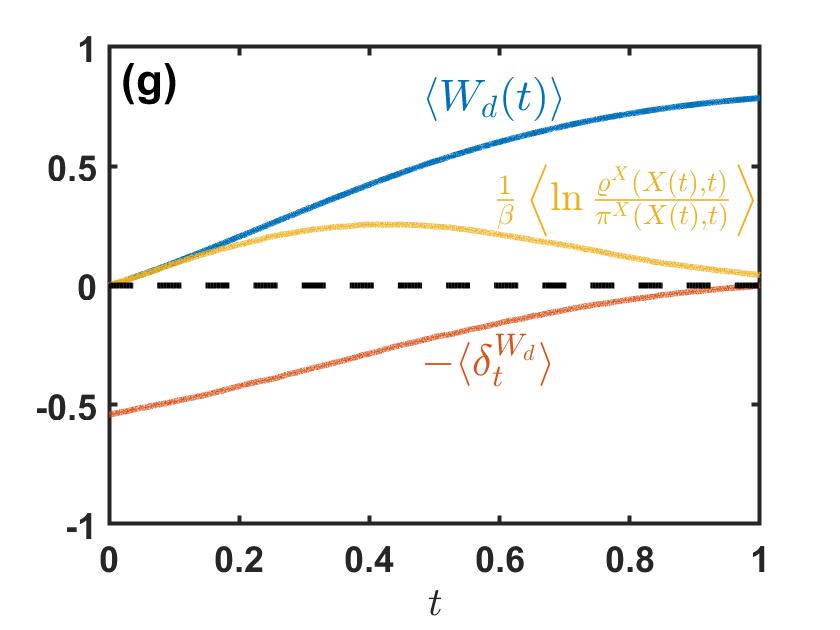}
	\end{subfigure}
	\caption{Numerical verification through a three-state jumping process(See maintext for details). (a) The entropy production $\sca{ S_\text{tot} (\tau) }$ (blue) and the corresponding compensation item $-\sca{ \delta^{S_\text{tot}}_\tau }$ (red) as functions of the threshold value $S_\text{th}$ for duration $T = 1$.
		(b) The free energy dissipation $\sca{ f_d (\tau) }$ (blue) and the corresponding compensation item $-\sca{ \delta^{f_d}_\tau }$ (red) as functions of the threshold value $f_\text{th}$ for duration $T = 1$.
		(c) The dissipative work $\sca{ W_d (\tau) }$ (blue) and the corresponding compensation item $-\sca{ \delta^{W_d}_\tau }$ (red) as functions of the threshold value $W_\text{th}$ for duration $T = 1$.
		(d),(e),(f) Test of the stopping-time fluctuation theorems \eqref{eq:thm_Stot}, \eqref{eq:thm_fd}, \eqref{eq:thm_Wd} with and without the compensated $\delta_t$.
		(g) When $\tau=t$ with probability $1$, the dissipative work $\sca{ W_d (t) }$ (blue), the corresponding compensation item $-\langle \delta^{W_d}_t \rangle$ (red), and the relative entropy $\sca{\ln \frac{\varrho^X (X(t), t)}{\pi^X (X(t), t)}}$ (yellow) as functions of $t$ for $0 \leqslant t \leqslant T$. \label{Discrete}}
\end{figure*}

Fig.~\hyperref[Discrete]{1(a-c)} shows the numerical results of $\sca{ S_\text{tot} (\tau) }$ versus $-\sca{ \delta^{S_\text{tot}}_\tau }$, $\sca{ f_d (\tau) }$ versus $-\sca{ \delta^{f_d}_\tau }$, and $\sca{ W_d (\tau) }$ versus $-\sca{ \delta^{W_d}_\tau }$, as functions of the threshold value. The initial distribution is set to be uniform among the three states in Fig.~\hyperref[Discrete]{1(a-b)}, and concentrated on the second state in Fig.~\hyperref[Discrete]{1(c)}. Fig.~\hyperref[Discrete]{1(d-f)} test the stopping-time fluctuation relations \eqref{eq:thm_Stot}, \eqref{eq:thm_fd} and \eqref{eq:thm_Wd}, with and without the compensated term $\delta_t$. 

In the special situation that $\tau=t$ with probability $1$, Fig.~\hyperref[Discrete]{1(g)} shows that the inequality \eqref{ineq2_cons_Wd} for $\sca{ W_d (t)}$ is not only stronger than the inequality $\sca{ W_d(t) } \geqslant - \langle \delta^{W_d}_t \rangle$, but also the traditional Jarzynski inequality $\sca{ W_d(t) } \geqslant 0$.

Another example is the stochastic dynamics of a colloidal particle with diffusion coefficient $D$ in a time-dependent potential $V(t)$. The dynamics obeys the Langevin equation
\begin{equation*}
	\frac{\dd X(t)}{\dd t} = - \frac{\partial V}{\partial x} (X (t), t) + \xi (t),
\end{equation*}
where $\xi$ is a Gaussian white noise with zero mean and autocorrelation $\sca{ \xi (t) \xi (t') } = 2D \delta (t - t')$. 

\begin{figure*}[htbp]
	\begin{subfigure}[b]{0.4\textwidth}
		\includegraphics[width=\textwidth]{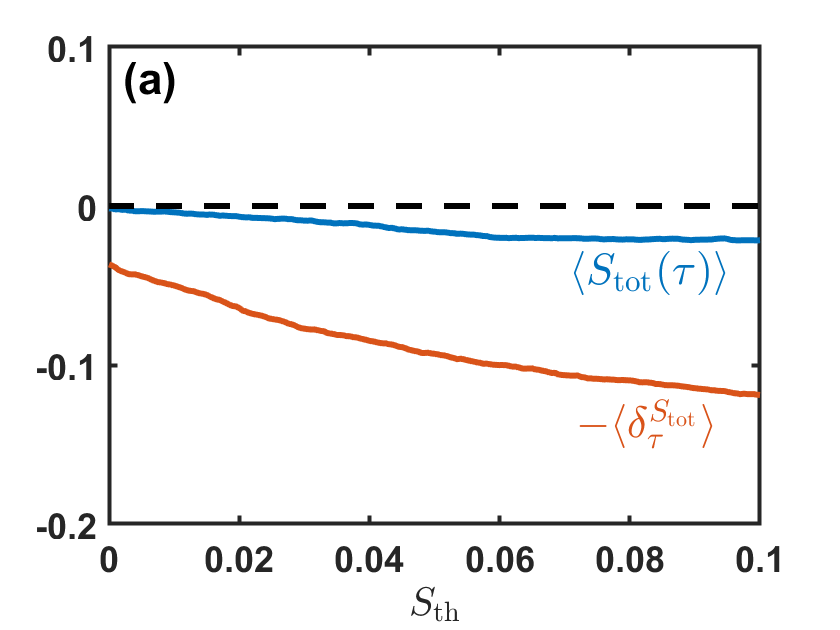}
	\end{subfigure}
	\begin{subfigure}[b]{0.4\textwidth}
		\includegraphics[width=\textwidth]{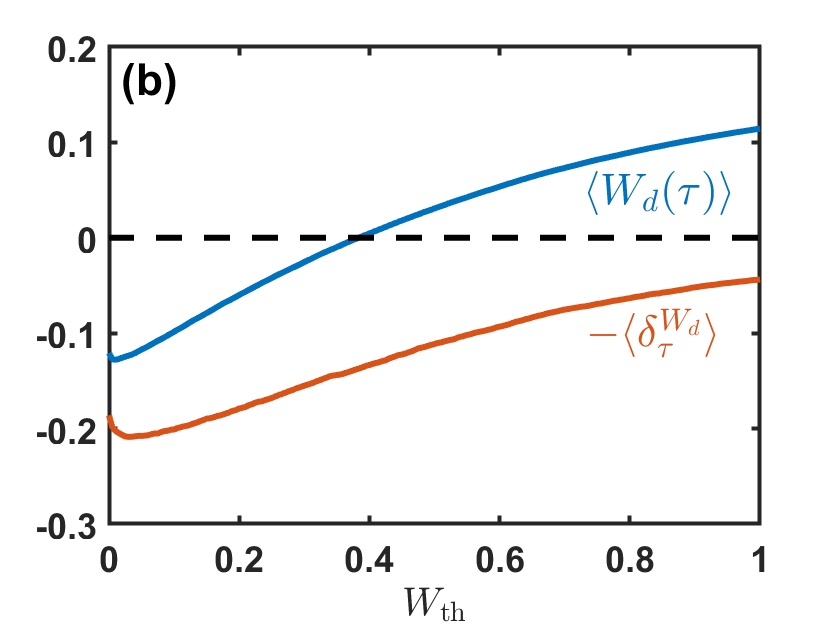}
	\end{subfigure}
	
	\begin{subfigure}[b]{0.4\textwidth}
		\includegraphics[width=\textwidth]{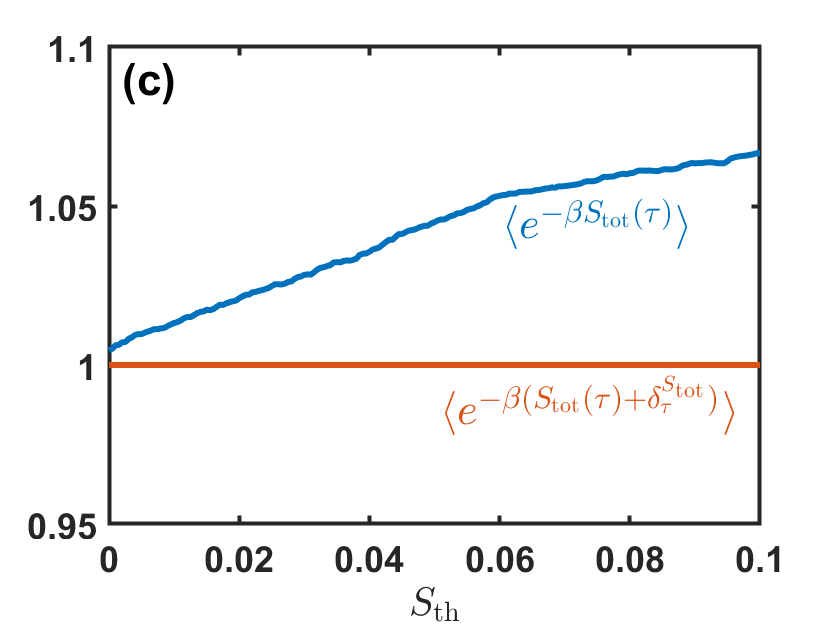}
	\end{subfigure}
	\begin{subfigure}[b]{0.4\textwidth}
		\includegraphics[width=\textwidth]{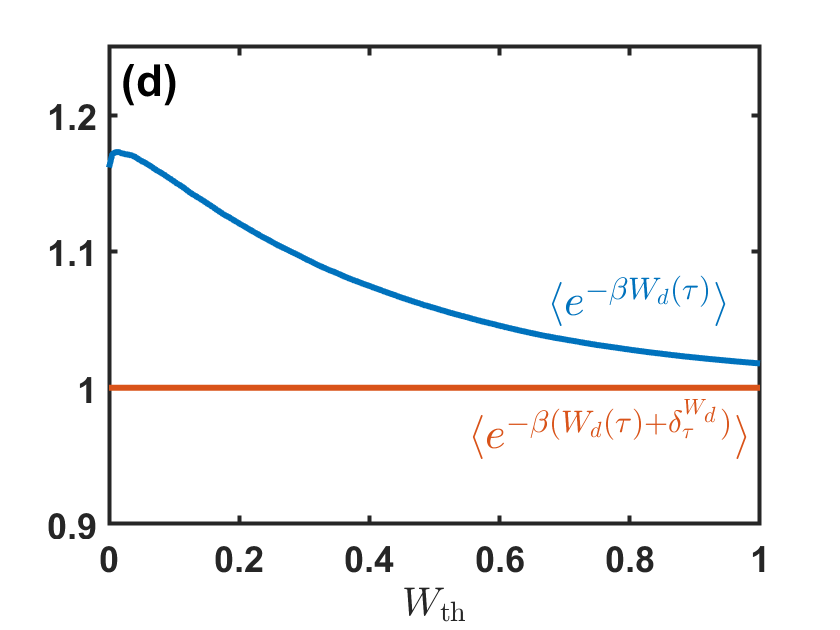}
	\end{subfigure}
	
	\begin{subfigure}[b]{0.4\textwidth}
		\includegraphics[width=\textwidth]{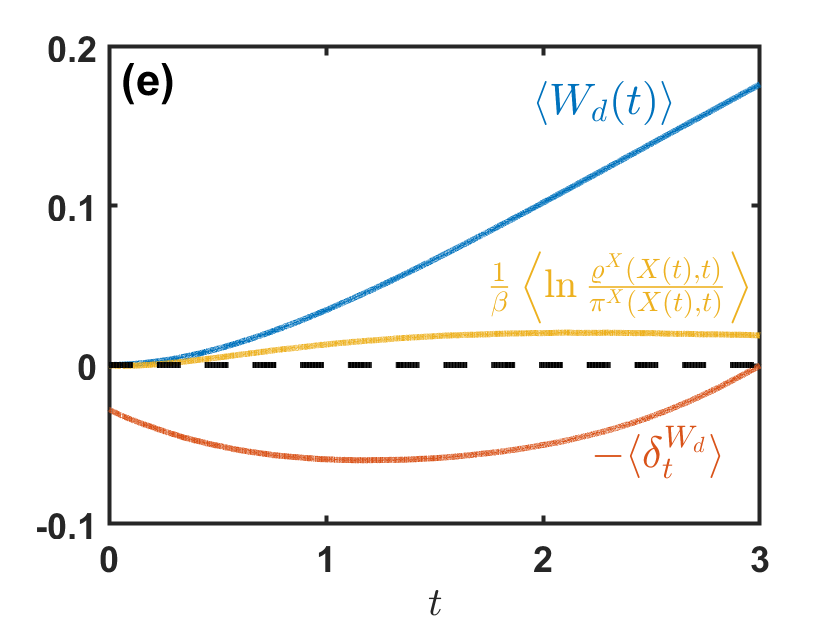}
	\end{subfigure}
	\caption{Numerical verification through a diffusion process(See maintext for details). (a) The entropy production $\sca{ S_\text{tot} (\tau) }$ (blue) and the corresponding compensation item $-\sca{ \delta^{S_\text{tot}}_\tau }$ (red) as functions of the threshold value $S_\text{th}$ for duration $T = 3$.
		(b) The dissipative work $\sca{ W_d (\tau) }$ (blue) and the corresponding compensation item $-\sca{ \delta^{W_d}_\tau }$ (red) as functions of the threshold value $W_\text{th}$ for duration $T = 3$.
		(c),(d) Test of the stopping-time fluctuation theorems \eqref{eq:thm_Stot}, \eqref{eq:thm_Wd} with and without $\delta_t$.
		(e) When $\tau=t$ with probability $1$, the dissipative work $\sca{ W_d (t) }$ (blue), the corresponding compensation item $-\langle \delta^{W_d}_t \rangle$ (red), and the relative entropy $\sca{\ln \frac{\varrho^X (X(t), t)}{\pi^X (X(t), t)}}$ (yellow) as functions of $t$ for $0 \leqslant t \leqslant T$.
		In this example, $V(x, t) = (t+4) (4x-t)^2 / 128$, $D = 1$.  \label{Continuous}}
\end{figure*}

In such a stochastic system, the housekeeping heat equals to zero and thus the entropy production $S_\text{tot} (t)$ coincides with the free energy dissipation $f_d (t)$. We follow the same stopping strategy as in the discrete model of Fig.~\hyperref[Discrete]{1}, and show the numerical results of $\sca{ S_\text{tot} (\tau) }$ versus $-\sca{ \delta^{S_\text{tot}}_\tau }$ in Fig.~\hyperref[Continuous]{2(a)} and $\sca{ W_d (\tau) }$ versus $-\sca{ \delta^{W_d}_\tau }$ in Fig.~\hyperref[Continuous]{2(b)} with $T = 3$. Fig.~\hyperref[Discrete]{2(c-d)} test the stopping-time fluctuation relations \eqref{eq:thm_Stot} and \eqref{eq:thm_Wd}, with and without the compensated term $\delta_t$.

In the special situation that $\tau=t$ with probability $1$, Fig.~\hyperref[Continuous]{2(e)} shows that the conclusion \eqref{ineq2_cons_Wd} for $\sca{ W_d (t)}$ is stronger than the inequality $\sca{ W_d(t) } \geqslant -\langle \delta^{W_d}_t \rangle$ and the Jarzynski inequality $\sca{ W_d(t) } \geqslant 0$.

In Fig.~\hyperref[Discrete]{1} and \hyperref[Continuous]{2}, the averaged thermodynamic functionals may be  negative under certain stopping strategy, but the general stopping-time fluctuation relations and related thermodynamic inequalities always hold.

{\em Derivation} First, we notice that
\begin{equation} \label{eq:pr1}
	\left\{ \frac{\PP^{\tilde{Z}^{T,t}} (\XX_{[0, t]} ) }{ \PP^X (X_{[0, t]}) } \right\}_{0 \leqslant t \leqslant T},
\end{equation}
is a martingale, where $\XX_{[0, t]} \equiv \{ X (t - s) \}_{0 \leqslant s \leqslant t}$ denotes the time reversal of $X_{[0, t]}$ in the duration $[0,t]$, and $\PP^{\tilde{Z}^{T,t}}(x_{[0,t]})$ denotes the probability of observing a given trajectory $x_{[0,t]}$ in $\{\tilde{Z}^{T,t}(s)=Z^T (s+T-t)\}_{0\leqslant s\leqslant t}$. The distribution of $Z^T(0)$ is $\tilde{\varrho}^{Z^{T}} (\cdot,0)$.

Since
\begin{equation*}
	\PP^X (X_{[0, T]})
	= \PP^X (X_{[0, T]} \vert X_{[0, t]}) \PP^X (X_{[0, t]}),
\end{equation*}
we have
\begin{eqnarray*}
	&&\mathrel{\phantom{=}} \sca{ \frac{\PP^{\tilde{Z}^{T,T}} (\XX_{[0, T]} ) }{ \PP^X (X_{[0, T]}) } \middle\vert X_{[0, t]} } \\
	&=& \sum_{X_{[t, T]}} \frac{\PP^{\tilde{Z}^{T,T}} (\XX_{[0, T]} ) }{ \PP^X (X_{[0, T]}) } \PP^X (X_{[0, T]} \vert X_{[0, t]}) \\
	&=& \sum_{X_{[t, T]}} \frac{\PP^{\tilde{Z}^{T,T}} (\XX_{[0, T]} ) }{ \PP^X (X_{[0, t]}) }. 
\end{eqnarray*}

For $0 \leqslant u \leqslant s \leqslant T$,  let $\XX_{[0, T]}(u,s)$ be the part of the trajectory $\XX_{[0, T]}$ in the duration $[u,s]$, then $X_{[t, T]}$ and $\XX_{[0, T]}(0,T-t)$ are exactly the time reversal of each other. Thus
\begin{align*}
	&\mathrel{\phantom{=}} \sum_{X_{[t, T]}} \PP^{\tilde{Z}^{T,T}} (\XX_{[0, T]} ) \\
	&= \sum_{\XX_{[0, T]}(0,T-t)} \PP^{\tilde{Z}^{T,T}} (\XX_{[0, T]} ) \\
	&= \PP^{\tilde{Z}^{T,T}} (\XX_{[0, T]} (T-t,T) ) \\
	&= \PP^{\tilde{Z}^{T,t}} (\XX_{[0, t]} ),
\end{align*}
in which the last equality comes from the definition of $\PP^{\tilde{Z}^{T,t}} (\XX_{[0, t]} )$. So
\begin{equation*}
	\sca{ \frac{\PP^{\tilde{Z}^{T,T}} (\XX_{[0, T]} ) }{ \PP^X (X_{[0, T]}) } \middle\vert X_{[0, t]} }
	= \frac{\PP^{\tilde{Z}^{T,t}} (\XX_{[0, t]} ) }{ \PP^X (X_{[0, t]}) },
\end{equation*}
which is exactly the definition of martingale for \eqref{eq:pr1}.

Second, we show that $\{ e^{-\beta (F_t+\delta_t)} \}_{0 \leqslant t \leqslant T}$ is exactly the martingale \eqref{eq:pr1}. By the definition of $F_t$, we have
\begin{equation} \label{eq:pr2}
	\frac{\PP^{\tilde{Z}^{T,t}} (\XX_{[0, t]} ) }{ \PP^X (X_{[0, t]}) } = \frac{\PP^{\tilde{Z}^{T,t}} (\XX_{[0, t]} ) }{ \PP^{Z^{t}} (\XX_{[0, t]} ) } e^{-\beta F_t}.
\end{equation}
Since
\begin{gather*}
	\PP^{\tilde{Z}^{T,t}} (\XX_{[0, t]} ) =\PP^{\tilde{Z}^{T,t}} ( \XX_{[0, t]} \vert \XX (0) ) \tilde{\varrho}^{Z^T} (X(t), T-t ), \displaybreak[3] \\
	\PP^{Z^t} (\XX_{[0, t]}) = \PP^{Z^t} ( \XX_{[0, t]} \vert \XX (0) ) \varrho^{Z^t} (X(t), 0 ),
\end{gather*}
and notice that $\{\tilde{Z}^{T,t} (s)\}_{0\leqslant s\leqslant t}$ and $\{Z^t (s)\}_{0 \leqslant s\leqslant t}$ are driven by the same protocol $\{ \tilde{\lambda}(t-s) \colon 0 \leqslant s \leqslant t \}$, we have
\begin{equation*}
	\PP^{\tilde{Z}^{T,t}} ( \XX_{[0, t]} \vert \XX (0) )
	= \PP^{Z^t} ( \XX_{[0, t]} \vert \XX (0) ),
\end{equation*}
which implies
\begin{equation} \label{eq:pr3}
	\frac{\PP^{\tilde{Z}^{T,t}} (\XX_{[0, t]} ) }{ \PP^{Z^t} (\XX_{[0, t]} ) }
	= \frac{\tilde{\varrho}^{Z^T} (X(t), T-t )}{\varrho^{Z^t} (X(t), 0 )}
	= e^{-\beta \delta_t}.
\end{equation}

Combining \eqref{eq:pr2} and \eqref{eq:pr3} shows that $\{ e^{-\beta (F_t+\delta_t)} \}_{0 \leqslant t \leqslant T}$ is exactly the martingale \eqref{eq:pr1}, then the general stopping-time fluctuation theorem \eqref{eq:thm} follows from the optional stopping theorem. 


When $\tau=t$ with probability $1$, we decompose
\begin{align*}
	&\mathrel{\phantom{=}} -\sca{\delta_t} \\
	&= \frac{1}{\beta} \sca{ \ln \frac{\tilde{\varrho}^{Z^T} (X(t), T - t )}{\varrho^{Z^t} (X(t), 0)} } \\
	&= \frac{1}{\beta} \sca{ \ln \frac{\varrho^X (X(t), t)}{\varrho^{Z^t} (X(t), 0)} } + \frac{1}{\beta} \sca{ \ln \frac{\tilde{\varrho}^{Z^T} (X(t), T - t )}{\varrho^X (X(t), t)}}.
\end{align*}
By Jensen's inequality, we know
\begin{gather*}
	\sca{ \ln \frac{\tilde{\varrho}^{Z^T} (X(t), T - t )}{\varrho^X (X(t), t)}}
	\leqslant \ln \sca{ \frac{\tilde{\varrho}^{Z^T} (X(t), T - t )}{\varrho^X (X(t), t)}}
	= 0.
\end{gather*}
Furthermore, for any given $t$, we can choose $\tilde{\varrho}^{Z^T} (\cdot, 0)$ such that $\tilde{\varrho}^{Z^T} (x, T-t) = \varrho^X (x, t)$, which leads to
\begin{equation*}
	\sup_{\tilde{\varrho}^{Z^T}}-\sca{ \delta_t }
	= \frac{1}{\beta} \sca{\ln \frac{\varrho^X (X(t), t)}{\varrho^{Z^t} (X(t), 0)}}
	\geqslant 0.
\end{equation*}

{\em Conclusion} 
In summary, our study contributes a general framework for understanding martingales constructed upon thermodynamic functionals. We have successfully derived and proven the stopping-time fluctuation theorems, accompanied by second-law-like inequalities for mean thermodynamic functionals stopped at stochastic times. Our results generalize the recent gambling strategy and stopping-time fluctuation theorems \cite{Manzano2021} to a very general setting. Our framework encompasses the general definition of thermodynamic functionals, accommodates various types of stochastic dynamics, and allows for arbitrary stopping strategies. The validity and applicability of our framework are supported by numerical verifications conducted in stochastic dynamics with both discrete and continuous states.

Furthermore, we highlight the significance of the additional degree of freedom introduced through the compensated term $\delta_t$, which leads to a strengthening of the inequality for dissipative work compared to the well-known Jarzynski inequality  when the stopping time is reduced to a deterministic one. Overall, our results provide novel insights, new interpretations, and improved bounds for the fundamental principles underlying the Second Law of Thermodynamics in the context of stochastic processes.

H. Ge is supported by NSFC 11971037 and T2225001.

\bibliography{bibliography}

\end{document}